\documentstyle[aps,preprint]{revtex}

\def\be{\begin{equation}}
\def\ee{\end{equation}}
\def\ba{\begin{eqnarray}}
\def\ea{\end{eqnarray}}

\def\Msolar{M_{\odot}}

\def\ga{\mathrel{\mathpalette\fun >}}
\def\fun#1#2{\lower3.6pt\vbox{\baselineskip0pt\lineskip.9pt
        \ialign{$\mathsurround=0pt#1\hfill##\hfil$\crcr#2\crcr\sim\crcr}}}

\begin{document}

\draft
\preprint{astro-ph/9907020 (Phys. Rev. D, in press, 2000)}
\date{February 18, 2000}
\title{Self-lensing of a Singular Isothermal Sphere}
\author{Yun Wang}
\address{Dept. of Physics,
		225 Nieuwland Science Hall, 
		University of Notre Dame, 
		Notre Dame, IN 46556-5670 }

\maketitle

\begin{abstract}
Many astrophysical systems can be approximated as isothermal spheres.
In an isothermal sphere, the ``foreground'' objects can act as lenses 
on ``background'' objects in the same distribution.
We study gravitational lensing by a singular isothermal sphere 
analytically. Our results may have interesting applications.

\end{abstract}


\newpage

\narrowtext

\section{Introduction}
Many astrophysical systems can be approximately described as isothermal
spheres, e.g. globular clusters, elliptical galaxies, clusters of
galaxies. They usually consist (at least partly) of  smaller 
``particles" which are more compact objects, like stars or galaxies. 
Seen from the outside, i.e. in projection, light from the ``background"
particles has to pass near the ``foreground" particles, and
hence it is affected by their gravitational attraction. This effect
of ``self-lensing" by a stellar disk has been studied in the past
\cite{Gould95}.
The self-lensing by an isothermal sphere is a simple, 
well defined problem which may have interesting applications. 
In this paper, we study the gravitational lensing of 
the constituents of a singular 
isothermal sphere (SIS) analytically. For simplicity of notation, 
we refer to the objects which constitute the SIS as ``stars''.

In \S2 we define some expressions which are used in the later sections.
In \S3 we give analytical expressions of the optical depths for
self-lensing by a SIS, the number of lensed stars, and the
duration distribution of lensing events.
In \S4, we derive the expressions for the number of lensed stars 
for self-lensing by a SIS, assuming an arbitrary distribution
in the luminosity of stars.
The special case of a power law distribution in the luminosity of stars 
is treated in \S5. In \S6, we apply our results to the centers
of the Milky Way and M87.
We give a brief summary in \S7.
Appendixes A and B contain results for a SIS lensing background
sources (located beyond the cut-off radius of the SIS) for comparison.

\section{Basic equations}

Here we give the basic equations and definitions which are used in
later sections.
A singular isothermal sphere (SIS) has the mass density
\be
\label{eq:rho}
\rho_L(r) = \rho_L(0)\, \left( \frac{r_c}{r} \right)^2 =
\frac{M_g}{4\pi r_g}\cdot \frac{1}{r^2}
\ee
at distance $r$ from the center, where $\rho_L(0)$ and $r_c$
are constants, and $M_g$ is the mass enclosed within radius $r_g$.
Let us assume that the SIS consists of only stars with mass $M_L$, 
with the number density $n_L(r)= \rho_L(r)/M_L$.

We consider a SIS with a cutoff radius $r_g$ and its center located 
at distance $r_0$ away from the observer.
The angular Einstein radius $\theta_E$ of an individual star is given by
\cite{SEF}
\be
\label{eq:theta_E}
\theta_E^2 = \frac{4GM_L}{c^2}\cdot \frac{D_{ds}}{D_d D_s},
\ee
where $D_s$ and $D_d$ are the distances of observer to source and lens, 
respectively, and $D_{ds}$ is the distance between the lens and the source.
Fig. 1 illustrates the geometry between observer, lenses 
and sources in the SIS. The following expressions will be useful:
\ba
\label{eq:defs}
r_d       &=& \sqrt{r_0^2+D_d^2-2r_0D_d \cos\theta}, \nonumber \\
r_s       &=& \sqrt{r_0^2+D_s^2-2r_0D_s \cos\theta}, \nonumber \\
r_1 	  &=& r_0\cos\theta- \sqrt{r_g^2-r_0^2\sin^2\theta}, \nonumber \\
r_2	  &=& r_0\cos\theta+ \sqrt{r_g^2-r_0^2\sin^2\theta}, \nonumber \\
\alpha    &\equiv& \frac{r_g}{r_0}, \nonumber \\              
\alpha_0 &\equiv& \sqrt{\alpha^2 -\sin^2\theta}
\ea
where $\theta$ is the angle between the line of sight to the source and
the line of sight to the center of the SIS;
$r_d$ and $r_s$ are the lens' and the source's distances from the center 
of the SIS respectively;
$r_1$ and $r_2$ are respectively the smallest and largest possible
distances from the observer to a star in the SIS for given $\theta$.

The optical depth for gravitational lensing for 
a source located in the direction $\theta$ and 
at a distance of $D_s$ from the observer is\footnote{Here we employ
	the definition for ``optical depth"  in the sense that
	it is the fractional area that is covered by the Einstein 
	circles of the lenses,  that means a minimum magnification
	of $\mu = 1.34$ is required, see Paczynski (1986)
\cite{Pa86}.}

\be
\tau(D_s, \theta)= \int^{D_{d,max}}_{D_{d,min}} dD_d\,
\pi \left[ D_d\, \theta_E\right]^2\, n_L(r_d), 
\ee
where $D_{d,max}$ and $D_{d,min}$ are the maximum and minimum distances
to a lens respectively.
The angular number density of lensed sources is
\be
\label{N_L(theta)def}
\frac{dN_L(\theta)}{d\theta}= 2\pi \sin\theta
\int_{D_{s,min}}^{D_{s,max}} dD_s\,
n_s(r_s)\, D_s^2 \, \tau(D_s, \theta) 
\ee
where $D_{s,max}$ and $D_{s,min}$ are the maximum and minimum distances
to a source respectively.
This should be compared with the angular number density of sources
\be
\frac{dN(\theta)}{d\theta}= 2\pi \sin\theta
\int_{D_{s,min}}^{D_{s,max}} dD_s\,
n_s(r_s)\, D_s^2  .
\ee

For a given direction $\theta$, the probability of finding a source which
is lensed is
\be
P(\theta) = \frac{dN_L(\theta)/d\theta}{dN(\theta)/d\theta}.
\ee
Again, ``lensed" here means the source lies inside the Einstein
radius of a foreground object.

\section{Analytical Results}

In this section, we calculate the optical depth for gravitational 
self-lensing by a singular isothermal sphere (SIS),
the number of lensed stars, and the duration distribution of lensing events.
Both lenses and sources are distributed throughout the SIS
(each constituent of the singular isothermal sphere can act both
as a lens and as a source).
The optical depth for gravitational lensing for a source located
at a distance of $D_s$ from the observer is
\ba
\label{eq:tau(Ds,theta)}
\tau(D_s, \theta)&= &\int^{D_s}_{r_1} dD_d\,
\pi \left[ D_d\, \theta_E\right]^2\, n_L(r_d)   \\
&=& \frac{GM_g}{c^2 r_g}\, \left\{ \frac{r_1}{D_s}-1
+\frac{1}{2} \left(1-\frac{2r_0 \cos\theta}{D_s} \right)\,
\ln\left[ \frac{ (D_s-r_0\cos\theta)^2+ r_0^2\sin^2\theta}
{r_g^2} \right] \right. \nonumber \\
&& \left.
+ \frac{D_s\cos\theta - r_0 \cos 2\theta}{D_s\sin\theta}\,
\left[\arctan\left(\frac{D_s-r_0\cos\theta}{r_0\sin\theta}\right)
+\arctan\left( \frac{\alpha_0}{\sin\theta}\right) \right] \right\}. \nonumber
\ea
For $\theta \ll 1$, $\tau(D_s,\theta)$ is dominated by
the last term in Eq.(\ref{eq:tau(Ds,theta)}); we find
\ba
\tau(D_s>r_0, \theta\ll 1) &\simeq& \frac{\pi G M_g}{c^2 r_g}\,
\frac{D_s-r_0}{D_s}\, \frac{1}{\theta},  \\
\tau(D_s<r_0, \theta\ll 1) &\simeq&  \frac{G M_g}{c^2 r_g}\, \left\{
\frac{1}{2} \left( \frac{2r_0}{D_s} -1 \right) \ln\left[ 
\frac{r_g^2}{(D_s-r_0)^2 +D_s r_0 \theta^2} \right] + 
\frac{1}{D_s} \left[ \frac{r_0-D_s}{\alpha}-r_g \right] -1 \right\}.\nonumber
\ea
Note that because the lens number density $n(r_d) \propto 1/r_d^2$,
for $D_s<r_0$ (source is closer to the observer than the center of the SIS),
the lens number density is finite everywhere,
$\tau(D_s,\theta)$ is roughly independent of $\theta$; 
while for $D_s>r_0$, the lens number density is infinite at 
$D_d=r_0$ and $\theta=0$,
$\tau(D_s,\theta)$ is roughly proportional to 1/$\theta$.
In Fig.2, we show the gravitational lensing optical depth in
the galactic center region due to self-lensing, 
assuming that the stellar cluster there is described by a SIS.
Fig.2 shows $\overline{\tau}(D_s,\theta)=\tau(D_s,\theta)\,
\left[GM_g/(c^2 r_g)\right]^{-1}$ for $r_0=8.5\,$kpc, as function of $D_s$
for $\theta=0.1''$, $1''$, and $10''$,
for SIS cutoff radius $r_g=100\,$pc (solid lines) 
and $1\,$kpc (dotted lines) respectively.

We assume that the distribution of sources in the SIS are also
of the same form as that of the lenses, i.e., $n_s(r_s) \propto
1/r_s^2$.
After integrating Eqs.(\ref{N_L(theta)def}) with the $\tau(D_s,\theta)$
from Eq.(\ref{eq:tau(Ds,theta)}), and taking $D_{s,min}=r_1$,
$D_{s,max}=r_2$ (see Fig.1), we find
\ba
\label{eq:I(theta)}
\frac{dN_L(\theta)}{d\theta}
&=& 8\pi n_s(0) r_c^2 r_0 \left(\frac{GM_g}{c^2 r_g}\right) 
 \left\{ \sin^2\theta\, 
\arctan\left(\frac{\alpha_0}{\sin\theta}\right) -
\alpha_0\sin\theta + \int^{\alpha_0/\sin\theta}_0
dx\, \frac{x \arctan x}{x^2+1} \right\}, \nonumber\\
\frac{dN(\theta)}{d\theta} &=&
4\pi n_s(0) r_c^2 r_0  \left[ \alpha_0\sin\theta +{\cos 2\theta}\,
\arctan\left(\frac{\alpha_0}{\sin\theta}\right) \right].
\ea
For $\theta \ll 1$, the angular densities of lensed sources $N_L$ and of 
all sources $N$ become
\ba
\label{eq:dNL}
\frac{dN_L(\theta)}{d\theta} &\simeq & 
8\pi n_s(0) r_c^2 r_0 \left(\frac{GM_g}{c^2 r_g}\right)
\left\{ \sin\theta^2\, \arctan\left(\frac{\alpha_0}{\sin\theta}\right) -
\alpha_0\sin\theta \right. \nonumber \\
\hskip 3.5cm && \left. +\ln\left(\frac{\alpha}{\sin\theta}\right)
\, \arctan\left(\frac{\alpha_0}{\sin\theta}\right)-
\frac{\pi}{2}\ln 2 \right\} \nonumber \\
& \simeq & 
4\pi^2 n_s(0) r_c^2 r_0 \left(\frac{GM_g}{c^2 r_g}\right)
\, \ln\left( \frac{\alpha}{2\theta} \right).\nonumber\\
\frac{dN(\theta)}{d\theta} &=& 2\pi^2 n_s(0) r_c^2 r_0  
\ea
The number densities of both sources and lensed sources increase rapidly 
as we go to small angles $\theta$.
The total number of lensed sources and of sources within angle $\theta_0$
of the line of sight of the SIS center are
\ba
N_L(\theta_0) &=& 16 \pi^3\, \frac{G\rho_L(0)}{c^2}\, n_s(0)\,r_c^4 
\,r_0\,\theta_0\, \left[\ln\left(\frac{\alpha}{2\theta_0}\right)+1\right], 
\nonumber\\
N(\theta_0) &=& 2\pi^2 \,n_s(0)\, r_c^2 \,r_0 \,\theta_0.
\label{eq:N_L}
\ea

The probability of finding a lensed source in the direction $\theta$ is 
\be
\label{eq:app. tau(theta)}
P(\theta) \simeq \frac{2GM_g}{c^2 r_g}\, \ln\left( \frac{\alpha}{2\theta} \right),
\hskip 1cm \theta \ll 1.
\ee
Note that $N_L(\theta)$ and $P(\theta)$ are not sensitive to $\alpha=r_g/r_0$,
the ratio of the SIS cutoff radius to the observer's distance to the SIS center.
Eqs.(\ref{eq:dNL})-(\ref{eq:app. tau(theta)}) are very good approximations 
for $\alpha \geq 10^{-2}$ and $\theta \leq 20''$.

The duration of a lensing event is given by 
\be
\label{eq:tDds}
\Delta t \sim \frac{r_0 \theta_E}{v_d} \propto \sqrt{D_{ds}}, 
\ee
where $v_d$ denotes the relative velocity of the lens with respect 
to the source.
In general, the lenses (and sources) have a distribution of velocities.

From the observational point of view, we would like to know
the number of lensing events which are shorter than a given time
threshold, say, the time duration of the observation. Only the
lensing events with time durations shorter than this threshold
can be observed with high certainty.
In our context, the number of lensing events with time duration
shorter than a given threshold, $t_c$, translates to the number 
of lensing events with lens-source separation $D_{ds}$ less than 
a length threshold $r_t(t_c,v_d)$, with
\be
r_t(t_c,v_d) = \frac{c^2}{4G M_L}\, t_c^2 v_d^2 \equiv A\,t_c^2\, v_d^2.
\ee
Within angle $\theta_0$ of the line of sight to the SIS center,
the number of lensing events with duration shorter than $t_c$ is
\be
\label{eq:NDt}
N_L(\Delta t <t_c, \theta_0)= \int_{-\infty}^{\infty} dv_d\, p(v_d)\,
N_L\left(D_{ds}\leq r_t(t_c,v_d), \theta_0\right),
\ee
where $p(v_d)$ is the probability distribution function of the
lens velocity $v_d$.

Within angle $\theta_0$ of the line of sight to the SIS center,
the number of lensing events with lens-source separation $D_{ds}$ less
than $r_t$ is given by
\ba
N_L(D_{ds}\leq r_t, \theta_0) &= &\int^{\theta_0}_0 2\pi \sin\theta\, d\theta
\left\{\int_{r_1}^{r_2} dD_s\, \int_{D_s-r_t}^{D_s} dD_d\, 
-\int_{r_1}^{r_1+r_t} dD_s\, \int_{D_s-r_t}^{r_1} dD_d\, \right\}
\nonumber \\
& & \hskip 1cm D_s^2 n_s(r_s) \, \pi (D_d\theta_E)^2 n_L(r_d)
\ea
If $r_0\theta_0 \ll r_t \ll r_g$, we find
\be
\label{eq:P(duration)}
N_L(D_{ds}\leq r_t, \theta_0) \simeq 16 \pi^3 \,
\frac{G \rho_L(0)}{c^2} \, n_s(0) r_c^4 
r_0 \theta_0 \left[ \ln\left(\frac{r_t}{2r_0\theta_0}\right)+1\right].
\ee

For any realistic distributions of the lens velocities $v_d$, say,
a Gaussian distribution, Eq.(\ref{eq:NDt}) has to be solved numerically. 
To obtain an order of magnitude estimate, we set the lens velocity 
to $v_d \sim v_0=\sqrt{4\pi G \rho_L(0) r_c^2}$.
Hence, the number of lensing events with time duration shorter than 
$t_c$, within angle $\theta_0$ is
\ba
N_L(\Delta t <t_c, \theta_0) &\sim& N_L\left(D_{ds}\leq r_t(t_c,v_0), 
\theta_0\right).
\nonumber \\
&\simeq & 16 \pi^3 \,\frac{G \rho_L(0)}{c^2} \, n_s(0) r_c^4 
r_0 \theta_0 \left[ \ln\left(\frac{A t_c^2 v_0^2}
{2r_0\theta_0}\right)+1\right],
\label{eq:NDta}
\ea
where the second approximation follows from Eq.(\ref{eq:P(duration)}),
and is only valid for 
$r_0\theta_0 \ll r_t(t_c,v_0)=A t_c^2 v_0^2 \ll r_g$, i.e., 
for lensing events with durations which are neither too long, 
nor too short.
Comparing Eq.(\ref{eq:P(duration)}) with Eq.(\ref{eq:N_L}), 
we find that 50\% of all lensing events have $D_{ds} \leq r_t$,
with $ r_0\theta_0 \ll r_t=0.86 \sqrt{r_g r_0\theta_0} \ll r_g$. 
This indicates that Eq.(\ref{eq:NDta}) is a useful approximation.

Note that we have assumed
that all lenses have the same mass for simplicity. 
The distribution in the lens mass, as well as the distribution
in lens/source velocities will contribute
to the broadening of the $\Delta t$ distribution.

\section{Number of lensed sources for an arbitrary luminosity function}

Observations are usually flux-limited, sources with apparent flux
below the flux limit of the observation will not be detected.
All known astrophysical sources have a distribution in their intrinsic
luminosities. The apparent flux of a source 
is determined by three factors:
the intrinsic luminosity of the source, the distance of the source,
and the magnification of the source due to gravitational lensing.
In this section, we examine the effect of all three factors 
on the number of lensed sources, allowing for an arbitrary distribution
in the intrinsic luminosity of the sources, since luminosity functions
are generally poorly known.

Let $\phi(M)$ be the luminosity function of the stars. 
The fraction of stars with intrinsic luminosity
greater than $f_0$ is
\be
\label{eq:Q0}
Q_0(f_0) = \frac{ \int_{-\infty}^{M_0} dM\,  \phi(M)}
{ \int_{-\infty}^{\infty} dM\,  \phi(M)},
\ee
$M_0$ is the absolute magnitude corresponding to $f_0$.
$Q_0(f_1)=1$ defines the low end of the luminosity distribution, $f_1$.
In general, the inclusion of a distribution in the luminosity of stars 
leads to a reduction in the expected number of lensed stars.

We write the apparent luminosity of a star with and without gravitational
lensing as
\be
f_L = \frac{C_1 \mu f_0}{D_s^2}; \hskip 1cm
f_{NL} = \frac{C_1 f_0}{D_s^2},
\ee
where $C_1$ is a constant, $\mu$ is the magnification,
$f_0$ is the intrinsic luminosity of the star, and $D_s$ is
the observer's distance to the star.
The fractions of stars with apparent luminosity greater than $f_{min}$ 
with and without gravitational lensing are
\be
Q_L(f_{min}) = Q_0\left(\frac{D_s^2 f_{min}}{C_1 \mu}\right),
\hskip 1cm
Q_{NL}(f_{min})  = Q_0\left(\frac{D_s^2 f_{min}}{C_1 }\right)
\ee
The number of sources with apparent luminosity greater than $f_{min}$
without gravitational lensing within angle $\theta_0$ from the line
of sight to the SIS center is
\be
N(\theta_0,f_{min}) = 2 \pi \int_0^{\theta_0} d\theta\, \sin\theta
\int_{D_{s,min}}^{D_{s,max}} dD_s\, D_s^2 \,n_s(r_s)\,Q_0(D_s^2 f_{min}/C_1).
\ee

For gravitational lensing by the SIS, the number of lensed sources with 
apparent luminosity greater than $f_{min}$ within angle $\theta_0$ is
\be
N_L(\theta_0,f_{min})= 2 \pi \int_0^{\theta_0} d\theta\, \sin\theta
\int_{D_{s,min}}^{D_{s,max}} dD_s\, D_s^2 \, n_s(r_s)\, 
\int_{f_1}^{\infty} df_0 \,
\left| \frac{d Q_0(f_0)}{d f_0} \right| \, \sigma(D_s,f_0, \mu),
\ee
where $\sigma(D_s,f_0, \mu)$ is the probability of
lensing with magnification greater than $\mu = f_{min}/f_{NL}
= f_{min} D_s^2/(f_0 C_1)$,
\be
\sigma(D_s,f_0, \mu) = \int_{D_{d,min}}^{D_s} dD_d \,
\pi \left[ D_d \theta_E \chi(\mu) \right]^2 \, n_L(r_d)
= \tau(D_s,\theta)\, \chi^2(\mu).
\ee
where $\chi(\mu)$ is the dimensionless angular position of the source 
relative to the optical axis (line connecting the observer and lens) in units
of $\theta_E$ for magnification $\mu$. Magnifications greater than $\mu$ 
require $\chi <\chi(\mu)$.

Gravitational lensing by a point mass lens leads to two images with
magnifications 
\be
\label{eq:mu}
\mu_{\pm}= \frac{1}{2} \left(1 \pm \frac{\chi^2+2}
{\chi \sqrt{\chi^2+4}}\right).
\ee
When the lens mass is of order $\Msolar$ (and the distances
involved are large), the resultant images
have micro arcsec scale separation and can not be resolved.
For unresolved images, we have
\be
\mu = \mu_{+}+\mu_{-} = \frac{\chi^2+2}{\chi \sqrt{\chi^2+4}}.
\ee
Hence we write
\be
\label{eq:chi2}
\chi^2(\mu)= \left\{
\begin{array}{ll} 2 \left( \frac{\mu}{\sqrt{\mu^2-1}} -1 \right), 
\hskip 1cm \mu>\mu_c, \\
1, \hskip 1cm \mu \leq \mu_c,
\end{array} \right.
\ee
where $\mu_c = 3/\sqrt{5}$.
This formula underestimates $\chi^2(\mu)$ for $\mu$ close to 1.

For resolved images, a lensed source can be detected if its brighter
image (with magnification $\mu_+$) has apparent luminosity greater than 
$f_{min}$. $\chi^2(\mu_+)$ is given by Eq.(\ref{eq:chi2}), with $\mu$ 
replaced by $2\mu_+ -1$ on the right hand side of the equation and  
$\mu_c=(3/\sqrt{5}+1)/2$. This would apply to the lensing of stars
by a central massive black hole; we mention it here
for completeness.

Using Eq.(\ref{eq:chi2}), we find
\be
N_L(\theta_0,f_{min})= 2 \pi \int_0^{\theta_0} d\theta\, \sin\theta
\int_{D_{s,min}}^{D_{s,max}} dD_s\, D_s^2 \, n_s(r_s)\, \tau(D_s,\theta)\,
Q(f_{min}, D_s),
\ee
with
\ba
Q(f_{min}, D_s) &= &\int_{f_1}^{\infty} df_0\, \left|
\frac{d Q_0(f_0)}{d f_0} \right|\, \chi^2(\mu) \\
&=& 2\, \int_{x_1}^{1/\mu_c} dx\, \frac{x}{(1-x^2)^{3/2}}\, Q_0(x)
+ 2\left( \frac{1}{\sqrt{1-x_1^2}} -1 \right), \nonumber
\ea
where  
\be
x= \frac{f_0 C_1}{f_{min} D_s^2}, \hskip 1cm
x_1= \frac{f_1C_1}{f_{min} D_s^2}. 
\ee
Note that $Q_0(x)=Q_0(f_0)$, and $f_1$ is defined by $Q_0(f_1)=1$.

The total number of lensed sources with unlensed apparent luminosity
greater than $f_{min}$ (i.e., they can be detected without the help
of lensing) is
\be
N_L(\theta_0, f_{min}, f_{NL}>f_{min}) = 
2 \pi \int_0^{\theta_0} d\theta\, \sin\theta
\int_{D_{s,min}}^{D_{s,max}}dD_s\, D_s^2
\, n_*(r_s)\,\tau(D_s,\theta)\, Q_0(D_s^2 f_{min}/C_1).
\ee
This gives us the number of sources which are detected due to gravitational 
lensing
\be
N_L(\theta_0, f_{min}, f_{NL}<f_{min}) =N_L(\theta_0, f_{min}) 
 -N_L(\theta_0, f_{min}, f_{NL}>f_{min}).
\ee

When the dimension of the source distribution is much smaller than
our distance to the SIS center (i.e., $r_g \ll r_0$ for self-lensing
of SIS), we can take 
$C_1=\overline{D_s}^2$, where $\overline{D_s}$ is the average distance
to the sources ($\overline{D_s}=r_0$ in the case of SIS self-lensing), 
then $x_1 \simeq f_1/f_{min}$, $Q_0(D_s^2 f_{min}/C_1) \simeq Q_0(f_{min})$,
and $Q(f_{min}, D_s) \simeq Q(f_{min})$.
The distribution in the intrinsic luminosities of the sources modifies
the total number of observed stars, and the total number of observed 
stars lensed by other stars (within angle $\theta_0$ of the line of sight
to the center of the SIS, and with apparent flux greater than $f_{min}$)
as follows
\ba
N(\theta_0,f_{min}) & \simeq & Q_0(f_{min})\, N(\theta_0),
\nonumber \\
N_L(\theta_0,f_{min}) &\simeq & Q(f_{min})\, N_L(\theta_0), 
\nonumber \\
N_L(\theta_0,f_{min},f_{NL}>f_{min}) &\simeq &
Q_0(f_{min})\,N_L(\theta_0), 
\nonumber \\
N_L(\theta_0,f_{min},f_{NL}<f_{min}) &\simeq &
\left[Q(f_{min})-Q_0(f_{min})\right]\,N_L(\theta_0), 
\ea
where $N(\theta_0)$ and $N_L(\theta_0)$ are given in the previous section.

\section{Number of lensed sources for a power-law luminosity function}

The simplest form for the fraction of stars with 
intrinsic luminosity greater than $f_0$ is
\be
\label{eq:Q0-pl}
Q_0(f_0) = \left\{ \begin{array}{ll}
	\left(\frac{f_0}{f_1}\right)^{-\beta}, \hskip 1cm \mbox{for }
	f_0>f_1; \\
	1, \hskip 1cm \mbox{for } f_0<f_1, \end{array} \right.
\ee
where $\beta$ is a positive constant.

The fractions of stars with apparent luminosity greater than $f$ 
with and without gravitational lensing are
\ba
Q_L(f) &=& \left\{ \begin{array}{ll}
\left( \frac{D_s^2 f}{C_1 \mu f_1} \right)^{-\beta}, \hskip 1cm
\mbox{if }\frac{D_s^2 f}{C_1 \mu} > f_1;\\
1, \hskip 1cm \mbox{else}.\end{array} \right. \nonumber \\
Q_{NL}(f) &= & \left\{\begin{array}{ll}
\left( \frac{D_s^2 f}{C_1 f_1} \right)^{-\beta}, \hskip 1cm
\mbox{if }\frac{D_s^2 f}{C_1} > f_1;\\
1, \hskip 1cm \mbox{else}.\end{array} \right.
\ea

The number of sources with apparent luminosity greater than $f_{min}$
without gravitational lensing within angle $\theta_0$ of 
$\theta=0$ is
\be
N(\theta_0,f_{min})  = 2 \pi \left(\frac{f_1}{f_{min}} \right)^{\beta}
\int_0^{\theta_0} d\theta\, \sin\theta\int_{D_{s,min}}^{D_{s,max}} dD_s\, 
D_s^2 \, \left(\frac{C_1}{D_s^2} \right)^{\beta} \,n_s(r_s),
\ee
where we have assumed $f_{min}\,D_s^2/C_1 >f_1$.

For gravitational lensing by the SIS, the number of lensed sources 
with apparent luminosity greater than $f_{min}$ within angle $\theta_0$ 
of $\theta=0$ is
\ba
N_L(\theta_0,f_{min}) &=&
2 \pi \left(\frac{f_1}{f_{min}} \right)^{\beta}
\int_0^{\theta_0} d\theta\, \sin\theta\int_{D_{s,min}}^{D_{s,max}} dD_s\, 
D_s^2 \, \left(\frac{C_1}{D_s^2} \right)^{\beta} \,n_s(r_s)
\nonumber \\
&&\hskip 1cm \cdot \tau(D_s,\theta)\,F(\beta, f_{min}, D_s),
\ea
with
\be
\label{eq:F}
F(\beta, f_{min}, D_s) = 2\beta \, \int_{x_1}^{1/\mu_c}
dx\, \frac{1}{x^{\beta+1}} \, \left( \frac{1}{\sqrt{1-x^2}}-1\right)
+ \mu_c^{\beta},
\ee
where $x_1=f_1C_1/(f_{min}D_s^2)$.
The number of lensed stars with unlensed flux greater than
$f_{min}$ is
\be
N_L(\theta_0,f_{min},f_{NL}>f_{min}) =
2 \pi \left(\frac{f_1}{f_{min}} \right)^{\beta}
\int_0^{\theta_0} d\theta\, \sin\theta\int_{D_{s,min}}^{D_{s,max}} dD_s\, 
D_s^2 \, \left(\frac{C_1}{D_s^2} \right)^{\beta} \,n_s(r_s)
\, \tau(D_s,\theta).
\ee
For $x_1 =f_1C_1/(f_{min}D_s^2)\ll 1$, we find
\be
F(\beta, f_{min}, D_s) = \left\{ \begin{array}{ll}
2.236-  x_1,
\hskip 1cm \beta=1; \\
1.677 - 2\, \ln x_1,
\hskip 1cm \beta=2.
\end{array} \right.
\ee

Note that for power-law $Q_0(f_0)$, $Q_0(f_{min})=(f_1/f_{min})^{\beta}$,
and $Q(f_{min})=(f_1/f_{min})^{\beta} F(\beta, f_{min}, D_s)$. 
Recall that 
$Q_0(f_{min})$ is the reduction factor in the total number of
observed stars (also the reduction factor in the total number of lensed stars
which can be observed without the help of gravitational lensing), 
$Q(f_{min})$ is the reduction
factor in the total number of observed stars lensed by other stars.
$\left[Q(f_{min})-Q_0(f_{min})\right]\,N_L(\theta_0)$
gives the number of stars observed due to gravitational lensing by the SIS.

\section{Applications to the centers of the Milky Way and M87}

An obvious application of our results is the self-lensing of
the star cluster near the Galactic center. Although the gravitational 
lensing near the Galactic center is dominated by the
central black hole \cite{Wardle92}, the number of 
lensed stars due to self-lensing of the star cluster near 
the Galactic center is only about an order of magnitude 
smaller than the number of stars lensed by the central black hole.
Gravitational lensing studies near the Galactic center can provide
independent evidence for the existence of the massive black hole at the
Galactic center, and constrain the stellar density near the Galactic 
center.

We take the distance to the Galactic center to be $r_0=8.5\,$kpc.
For the density profile of the SIS (see Eq.(\ref{eq:rho})), we take
$\rho_L(0)=4\times 10^6 \Msolar\,\mbox{pc}^{-3}$,  
and $r_c=0.17\,$pc.
In a cone with angular radius 
$\theta_0$ centered around our line of sight to the Galactic
center, without considering the distribution in the luminosity of the stars,
the number of lensed stars due to self-lensing of the star-cluster is 
[see Eq.(\ref{eq:N_L})]
\ba
\label{eq:N_L,GC}
N_L^{GC}(\theta_0)&=& 160 \pi^3 \,\frac{G\rho_L^2(0) r_c^4}{M_L c^2} \,r_0
\theta_0\, \gamma(\theta_0,r_g)\\
&=&0.1345 \, \gamma(\theta_0, r_g)\,
\left(\frac{\theta_0}{1''}\right)\, \left(\frac{\Msolar}{M_L}\right),
\ea
where $r_g$ is the cutoff radius of the SIS, and
$\gamma(\theta_0, r_g) \equiv \ln[r_g/(2 r_0\theta_0)+1]/10$.
Taking $\theta_0=1'' $, for $r_g=100$pc, 1kpc, and 10kpc,
$\gamma=0.81$, 1.04, and 1.27 respectively.

When we include the distribution in the luminosity of the stars
by using the empirical luminosity function from Mamon \& Soneira (1982)
\cite{MaSo82}, 
the number of lensed stars is reduced by approximately one order of 
magnitude from that given by Eq.(\ref{eq:N_L,GC}), for limiting
K magnitude of 21 and an extinction of 3 magnitudes to the Galactic center.

The duration of a lensing event is given by 
\be
\Delta t \sim \frac{r_0 \theta_E}{v_d} \sim
2 \, \mbox{days} \left(\frac{M_L}{\Msolar}\right)^{1/2}
\, \left( \frac{D_{ds}}{\mbox{pc}}\right)^{1/2}
\ee
where $v_d= \sqrt{4\pi G \rho_L(0) r_c^2} \simeq 80\,$km/s
is the mean velocity of the lens, $\theta_E$ is the angular Einstein 
radius given by Eq.(\ref{eq:theta_E}), and $D_{ds}$ is the separation
between the source and the lens.

The number of lensing events with lens-source separation $D_{ds}$ less
than $r_t$ within angle $\theta_0$ of the line of sight to the SIS center,
$N_L(D_{ds}\leq r_t, \theta_0)$, is given by Eq.(\ref{eq:P(duration)}).
50\% of all lensing events have $D_{ds} \leq 0.86 \sqrt{r_g r_0\theta_0}
\ll r_g$ (\S 3), i.e., roughly 50\% of all lensing events have duration 
shorter than 
\be
\Delta t \sim 2.64 \, \mbox{days} \left(\frac{\theta_0}{1''}\right)^{1/4}\, \left(\frac{r_g}{100\mbox{pc}}\right)^{1/4}\,
\left(\frac{M_L}{\Msolar}\right)^{1/2}.
\ee

The observation of the self lensing of stars near the Galactic center 
would require substantial observational resources, and should only be
a by-product of ultra high resolution proper motion studies of
stars near the Galactic center. Another by-product will be
observing the lensing of the same stars due to the 
black hole at the Galactic center. Note that the lensing events due to 
the central black hole last about three order of magnitude longer than the
self-lensing events. For a given observation with finite duration
and dense time sampling (say, the detailed and accurate monitoring of 
stellar motions near the Galactic center for two years), the smaller 
number of self-lensing events is 
balanced by their shorter time durations in favor of their detection.

Note that the number of lensed events increases with both typical
lens velocity $v_d= \sqrt{4\pi G \rho_L(0) r_c^2}$, and our distance
to the source $r_0$ [see Eqs.(\ref{eq:N_L}) and (\ref{eq:N_L,GC})].
For galaxies which are more distant from us or/and have higher central mass density, the number of lensed stars increases.
Our distance to M87 is about 15$\,$Mpc. 
Within 1 arcsecond (about 73$\,$pc) of the center of M87, 
the velocity dispersions of stars are about
400$\,$km/s \cite{Dehnen}.
This implies that M87 has a central stellar density at least as
large as that of the Milky Way.
For M87, 
$\gamma(\theta_0, r_g)=\ln[r_g/(2 r_0\theta_0)+1]/10
\sim 0.1$. Therefore 
\be
N_L^{M87}(\theta_0) \ga 176.5\,N_L^{GC}(\theta_0) 
\left(\frac{\theta_0}{1''}\right)\, \left(\frac{\Msolar}{M_L}\right)
\sim 24 \,\left(\frac{\theta_0}{1''}\right)\, \left(\frac{\Msolar}{M_L}\right).
\ee
If a large enough number of self-lensed stars are observed near the
center of M87, they can be used to constrain the model for
the central mass distribution of M87.

\section{Summary}

We have studied the self-lensing of a singular isothermal sphere (SIS)
analytically. We have derived simple analytical formulas for
the optical depth, the number of lensed stars, and the 
duration distribution of lensing events. 
We have also derived the expressions for the number of lensed stars 
assuming an arbitrary distribution
in the luminosity of stars, and we have considered
the special case of a power law distribution in the luminosity of stars.
Application of our results to the gravitational lensing by the star
cluster near the Galactic Center
gives potentially observable number of lensing events.
High resolution monitoring of the central region of M87 may
yield dozens of self-lensed stars which can be used to
contrain the model for the central mass distribution of M87.

For comparison, we give analytical expressions of the optical depth for
a SIS lensing background sources (located 
beyond the cut-off radius of the SIS) in Appendix A, 
and the application of these expressions to the gravitational
lensing of distant sources by the star cluster 
near the Galactic Center in Appendix B.

Since many astrophysical systems can be approximated as
isothermal spheres, it is possible that some of them can have
significant self-lensing;
our results should be useful in studying gravitational
lensing in these systems.

\acknowledgments It is a pleasure for me to thank Ed Turner 
for suggesting this calculation;
Joachim Wambsganss for providing
Figure 1, helpful discussions, and careful readings of the
manuscript; and the referee for helpful suggestions.

\vskip .2in

\appendix{\bf{Appendix A. SIS lensing a background source}}

For a source located beyond the cutoff radius of the SIS 
(on the other side of the SIS, away from the observer) and is lensed 
by the SIS, $D_s > r_2$ (see Fig.1), the optical depth is given by
\ba
\tau(D_s, \theta)&=&\int^{r_2}_{r_1} dD_d\,
\pi \left[ D_d\, \theta_E\right]^2\, n_L(r_d),  \\
&=& \frac{2GM_g}{c^2 r_g}\, \left\{\frac{D_s\cos\theta 
- r_0 \cos 2\theta}{D_s\sin\theta}\,
\arctan\left( \frac{\alpha_0}{\sin\theta} \right) 
-\frac{ r_0}{D_s} \, \alpha_0\right\}. \nonumber
\ea
For a source far behind the SIS, $D_s \gg r_2$, we find
\be
\tau(D_s\gg r_2, \theta)= \frac{2GM_g}{c^2 r_g}\,\cot\theta \,
\arctan\left( \frac{\alpha_0}{\sin\theta} \right),
\ee
which is independent of $D_s$; this is
consistent with the trend shown by Fig.2.
For $\theta \ll 1$, we find
\be
\tau(D_s, \theta\ll 1) \simeq \frac{\pi G M_g}{c^2 r_g}\,
\frac{D_s-r_0}{D_s}\, \frac{1}{\theta}.
\ee

Let us assume that the cluster of sources is located behind the SIS,
at a mean distance of $\overline{D_s}$ from the center of the SIS,
and that the radius of the distribution along the line of sight is
$r_s$. If $\overline{D_s} \gg r_s$, we can take $n_s(D_s)\simeq 
n_s(\overline{D_s})$. The angular densities of lensed sources and
of all sources become
\ba
\frac{dN_L(\theta)}{d\theta}
&=& 8\pi n_s(\overline{D_s}) \,\overline{D_s} \,r_s r_0 
\left(\frac{GM_g}{c^2 r_g}\right)  \left\{  
\left[ \frac{\overline{D_s} }{r_0} \cos\theta  -\cos 2\theta\right]
\arctan \left( \frac{\alpha_0}{\sin\theta}\right)
-\alpha_0 \sin\theta\right\}, \nonumber\\
\frac{dN(\theta)}{d\theta} &=&
4\pi n_s(\overline{D_s}) \, \overline{D_s}^2\, r_s \sin\theta.
\ea
For $\theta \ll 1$,
the angular densities of lensed sources and of all sources become
\ba
\frac{dN_L(\theta)}{d\theta} &\simeq & 
4\pi^2 n_s(\overline{D_s}) \,\overline{D_s} \,(\overline{D_s}-r_0)\,
 r_s \left(\frac{GM_g}{c^2 r_g}\right), \nonumber\\
\frac{dN(\theta)}{d\theta} &=&  4\pi n_s(\overline{D_s}) \,\overline{D_s}^2 
\, r_s \theta.
\ea

The probability of finding a lensed source in the direction $\theta$ is 
(see Eq.(7))
\be
P(\theta) \simeq \frac{GM_g}{c^2 r_g}\, \frac{\overline{D_s}-r_0}
{\overline{D_s}}\, \frac{1}{\theta},
\hskip 1cm \theta \ll 1.
\ee

Note that although the optical depth in the case of SIS lensing a 
background source has the same small $\theta$ limit as the optical
depth in the case of SIS lensing itself, the angular number density
of lensed sources increases more rapidly as $\theta$ decreases
in the case of SIS self-lensing due to the
increase in the number of sources.

\appendix{{\bf Appendix B. Lensing of distant sources by the star cluster 
near the Galactic Center}}

Again we take SIS density profile for the star cluster near the Galactic 
center, with $\rho_L(0)=4\times 10^6 \Msolar\,\mbox{pc}^{-3}$
and $r_c=0.17\,$pc.
For a cluster of sources located on the other side of the Galaxy,
we find
\be
N_L(\theta_0)=0.2 \,\left(\frac{\theta_0}{1''}\right)\, 
\left( \frac{n_s(\overline{D_s})}{1\, \mbox{pc}^{-3}}\right)
\left(\frac{\overline{D_s}}{17\,\mbox{kpc}}\right)
\left(\frac{\overline{D_s}-r_0}{8.5\,\mbox{kpc}}\right)
\left(\frac{r_s}{100\,\mbox{pc}}\right).
\ee
This number should probably decrease by more than an order of magnitude
when the luminosity distribution of stars and dust extinction are
taken into account.

The duration of a lensing event is given by 
\be
\Delta t \sim \frac{\overline{D_s} \theta_E}{v_s} \sim
8 \, \mbox{months} \left(\frac{M_L}{\Msolar}\right)^{1/2}
\, \left( \frac{80\,\mbox{km/s}}{v_s} \right)
\left( \frac{\overline{D_{s}}}{17 \mbox{kpc}}\right)^{1/2}
\left( \frac{\overline{D_{s}}-r_0}{8.5 \mbox{kpc}}\right)^{1/2}
\ee
The durations of lensing events are clustered around the typical value
because all the sources are at about the same distance behind the lens.


\clearpage

\begin{figure}
\caption[fig1.eps]
{Cartoon illustrating the geometry between observer (O), lenses (L)
and sources (S) in the singular isothermal sphere (SIS, represented
by the circle centered at C with radius $r_g$).
$\theta$ is the angle between the line of sight to the source (OS) and
the line of sight to the center of the SIS (OC);
the distances $r_d=$LC, $r_s=$SC,
$D_d=$OL, $D_{ds}=$LS, $D_s=$OS.
$r_1$ and $r_2$ are, respectively, the smallest and largest possible
distances from the observer (O) to an object in the SIS for given $\theta$.}
\end{figure}

\begin{figure}
\caption[fig2.eps]
{Optical depth for self lensing by a singular isothermal sphere (SIS)
at the Galactic center (at a distance of 8.5 kpc),
$\overline{\tau}(D_s,\theta)=\tau(D_s,\theta)\,
\left[GM_g/(c^2 r_g)\right]^{-1}$,
as function of the distance to the source $D_s$,
for source direction angle (measured from the line of sight to the 
Galactic center) $\theta=0.1''$, $1''$, and $10''$, and 
for two values of the cut-off radius of the SIS: 
$r_g=100\,$pc (solid lines) and $1\,$kpc (dotted lines).}
\end{figure}

\clearpage

\setcounter{figure}{0}

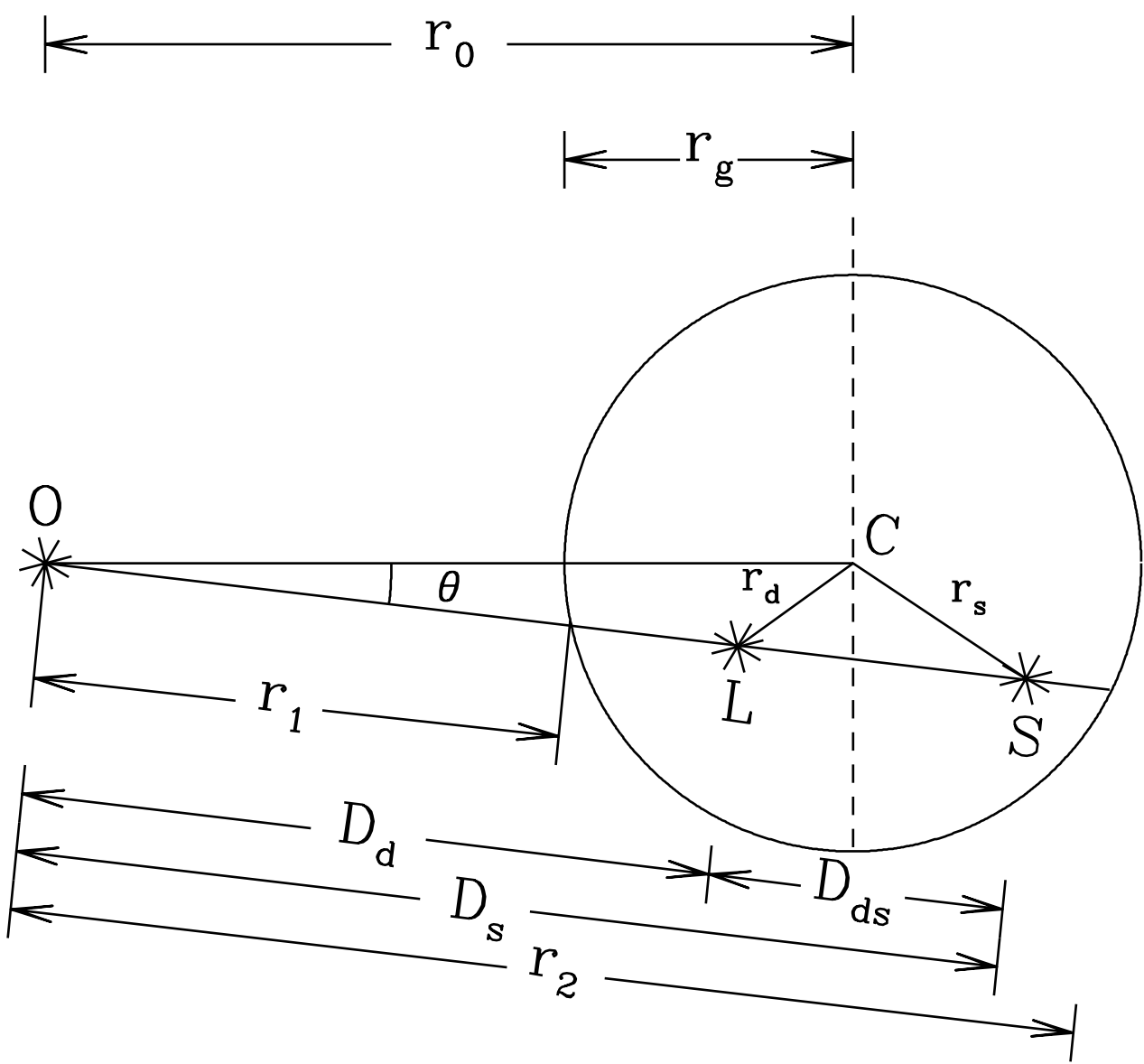
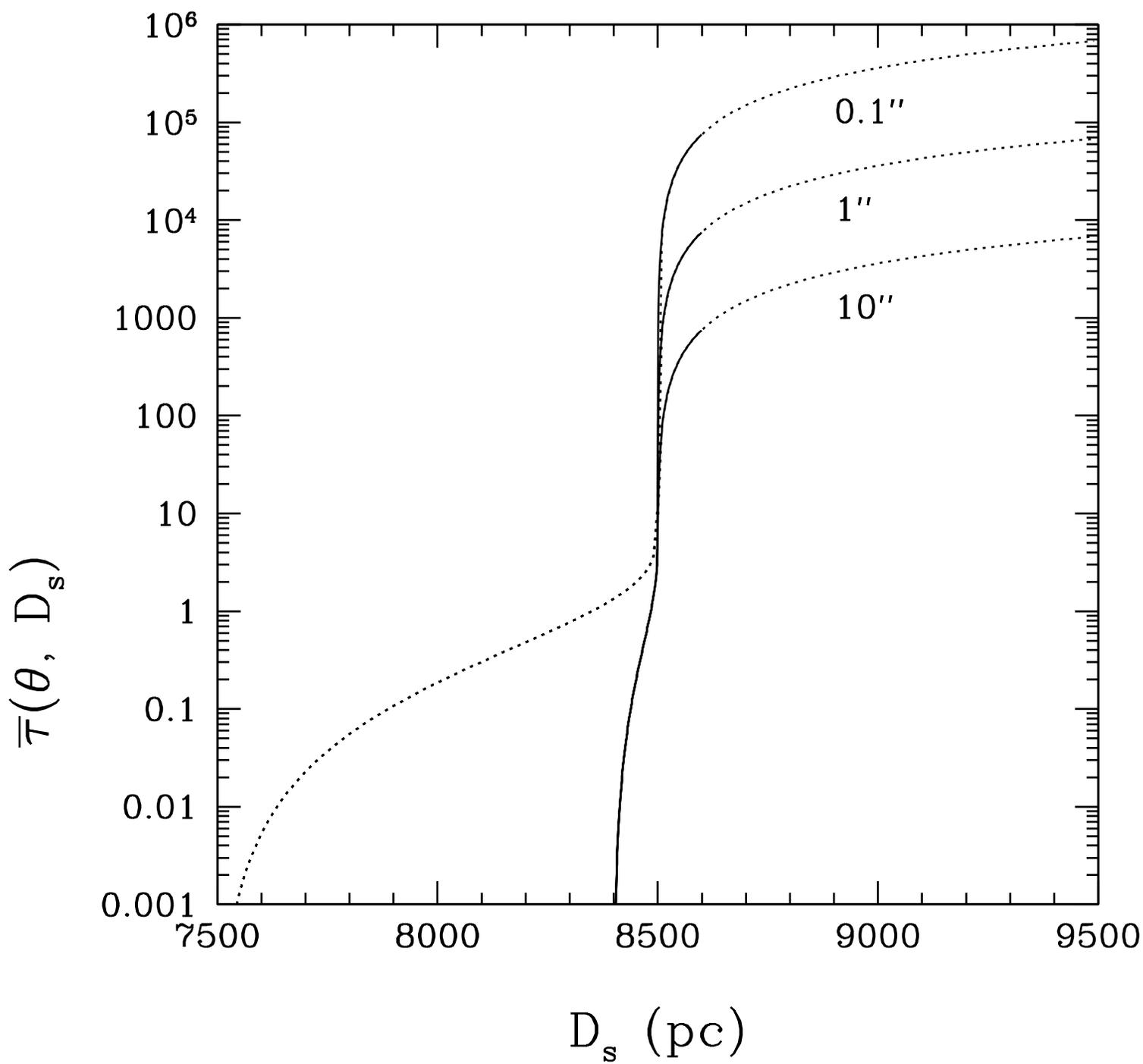
\end{document}